\documentclass[12pt,a4paper,DIV12]{scrartcl}
\usepackage[utf8x]{inputenc}
\usepackage[T1]{fontenc}
\usepackage{lmodern}
\usepackage[british]{babel}
\usepackage{graphicx}
\usepackage{hyperref}
\usepackage{color}
\usepackage{amsmath}
\usepackage{amssymb}
\usepackage{amsfonts}
\usepackage{authblk}
\newcommand*\samethanks[1][\value{footnote}]{\footnotemark[#1]}
\newcommand{\arxiv}[2]{[arXiv:\,\href{http://arxiv.org/abs/#1}{\texttt{#1}} [\texttt{#2}]]}
\newcommand{\arxivold}[1]{[arXiv:\,\href{http://arxiv.org/abs/#1}{\texttt{#1}}\,]}
\newcommand{\Tr}{\ensuremath{\mathrm{Tr}}}
\newcommand{\api}{\text{a--}\pi}
\title{%
   {\vspace{-20mm}\normalsize
    \hfill\parbox[b][30mm][t]{35mm}{\textmd{MS-TP-18-08}}}\\[-18mm]
Analysis of Ward identities in supersymmetric Yang-Mills theory
\vspace*{2mm}}
\author[1,2]{Sajid Ali%
\thanks{\{sajid.ali,h.gerber,munsteg,scior\}@uni-muenster.de}}
\author[3,1]{Georg Bergner\thanks{georg.bergner@uni-jena.de}}
\author[1]{Henning Gerber\samethanks[1]}
\author[4]{\newline Istvan Montvay\thanks{montvay@mail.desy.de}}
\author[1]{Gernot M\"unster\samethanks[1]}
\author[5]{Stefano Piemonte\thanks{stefano.piemonte@ur.de}}
\author[1]{Philipp Scior\samethanks[1]}
\affil[1]{University of M\"unster, Institute for Theoretical Physics,
Wilhelm-Klemm-Str.~9, D-48149 M\"unster, Germany}
\affil[2]{Government College University Lahore, Department of Physics,
Lahore 54000, Pakistan}
\affil[3]{University of Jena, Institute for Theoretical Physics,
Max-Wien-Platz 1, D-07743 Jena, Germany}
\affil[4]{Deutsches Elektronen-Synchrotron DESY,
Notkestr.~85, D-22607 Hamburg, Germany}
\affil[5]{University of Regensburg, Institute for Theoretical Physics,
Universit\"atsstr.~31, D-93040 Regensburg, Germany}
\date{February 27, 2018\\(revised April 30, 2018)}
\begin{document}
\maketitle

\newpage

\begin{abstract}
\noindent
\textbf{\textsf{Abstract:}}

In numerical investigations of supersymmetric Yang-Mills theory on a
lattice, the supersymmetric Ward identities are valuable for finding the
critical value of the hopping parameter and for examining the size of
supersymmetry breaking by the lattice discretisation. In this article we
present an improved method for the numerical analysis of supersymmetric Ward
identities, which takes into account the correlations between the various
observables involved. We present the first complete
analysis of supersymmetric Ward identities in $\mathcal{N}=1$ supersymmetric
Yang-Mills theory with gauge group SU(3). The results indicate that lattice
artefacts scale to zero as $O(a^2)$ towards the continuum limit in agreement
with theoretical expectations.
\end{abstract}

%%%%%%%%%%%%%%%%%%%%%%%%%%%%%%%%%%%%%%%%%%%%%%%%%%%%%%%%%%%%%%%%%%%%%%%%%%%%%%%%
\section{Introduction}

Ward identities are the key instruments for studying symmetries in quantum
field theory. They represent the quantum counterparts to Noether's
theorem, expressing the realisation of a classical symmetry at the quantum
level in terms of relations between Green's functions. They also allow to
characterise sources of explicit symmetry breaking. In the case of theories
that are regularised non-perturbatively by means of a space-time lattice,
Ward identities are a useful tool for the investigation of lattice
artefacts, which are related to the breaking of symmetries. In
lattice QCD, for example, chiral Ward identities in the form of the PCAC
relation are being used to quantify the breaking of chiral symmetry by the
lattice discretisation, and thereby to control the approach to the continuum
limit \cite{Luscher:1996sc}.

For supersymmetric (SUSY) theories the corresponding relations are the
supersymmetric Ward identities. In the context of numerical investigations
of supersymmetric Yang-Mills theory on a lattice, SUSY Ward identities are
being employed for a two-fold purpose \cite{Farchioni:2001wx}. First, in
numerical simulations using Wilson fermions a gluino mass is introduced,
which breaks supersymmetry softly. With the help of SUSY Ward identities
the parameters of the model can be tuned such that an extrapolation to
vanishing gluino mass is possible. Second, the discretisation on a lattice
generically breaks supersymmetry \cite{Bergner:2009vg}, leading to lattice
artefacts of order $a$ in the lattice spacing. By means of SUSY Ward
identities it can be checked if lattice artefacts are small enough for an
extrapolation to the continuum limit.

Our collaboration has employed SUSY Ward identities in previous
investigations of $\mathcal{N}=1$ supersymmetric Yang-Mills theory with
gauge group SU(2); for recent result see \cite{Bergner:2015adz}. In the
analysis of SUSY Ward identities, following the methods introduced in
\cite{Farchioni:2001wx}, the correlations between the various quantities
entering the calculation are, however, not being taken into account.
Therefore, for our present studies with gauge group SU(3) we developed a
method, based on a generalised least squares fit, that incorporates these
correlations. In this article we describe the method and present the results
of the first complete analysis of SUSY Ward identities for supersymmetric
Yang-Mills theory with gauge group SU(3).

%%%%%%%%%%%%%%%%%%%%%%%%%%%%%%%%%%%%%%%%%%%%%%%%%%%%%%%%%%%%%%%%%%%%%%%%%%%%%%%%
\section{Supersymmetric Ward identities on the lattice}

The $\mathcal{N}=1$ supersymmetric Yang-Mills (SYM) theory is the
supersymmetric extension of Yang-Mills theory with gauge group SU($N_c$). It
represents the simplest field theory with supersymmetry and local gauge
invariance. In the present investigations of our collaboration
\cite{Ali:2016zke} we are focussing on gauge group SU(3). SYM theory
describes the carriers of gauge interactions, the ``gluons'', together with
their superpartners, the ``gluinos'', forming a massless vector
supermultiplet. The gluons are represented by the non-Abelian gauge field
$A^{a}_{\mu} (x)$, $a = 1, \dots , N^{2}_{c} - 1$. The gluinos are massless
Majorana fermions, described by the gluino field $\lambda^{a} (x)$ obeying
the Majorana condition $\bar{\lambda} = \lambda^{T} C$ with the charge
conjugation matrix $C$, thus being their own antiparticles. Gluinos
transform under the adjoint representation of the gauge group, so that the
gauge covariant derivative is given by $(\mathcal{D}_{\mu} \lambda)^{a} =
\partial_{\mu} \lambda^{a} + g\,f_{abc} A^{b}_{\mu} \lambda^{c}$. In the
Euclidean continuum the (on-shell) Lagrangian of the theory, where auxiliary 
fields have been integrated out, is
\begin{equation}
\mathcal{L} = 
\frac{1}{4} F_{\mu\nu}^{a} F_{\mu\nu}^{a} +
\frac{1}{2} \bar{\lambda}^{a} \gamma_{\mu} (\mathcal{D}_{\mu}
\lambda)^{a}\,,
\end{equation}
where $F_{\mu\nu}^{a}$ is the non-Abelian field strength. Adding a gluino
mass term $(m_{0}/2)\ \bar{\lambda}^{a} \lambda^{a}$, which is
necessary in view of the numerical simulations, breaks supersymmetry softly.

Infinitesimal supersymmetry transformations, that leave the action of the
massless theory invariant, are given by
\begin{align}
\delta A_{\mu}^{a}(x)
&= 2 \bar{\lambda}^{a}(x) \gamma_{\mu} \epsilon,\nonumber\\
\delta \lambda^{a}(x)
&= + \sigma_{\mu\nu} F_{\mu\nu}^{a}(x) \epsilon,\\ 
\delta \bar{\lambda}^{a}(x)
&= - \bar{\epsilon} \sigma_{\mu\nu} F_{\mu\nu}^{a}(x),
\nonumber
\end{align}
where $\sigma_{\mu\nu} = (1/2) [ \gamma_{\mu}, \gamma_{\nu} ]$, and the
parameter $\epsilon$ is a Grassmann valued spinor. Noether's theorem,
applied to the classical theory, yields a supercurrent \cite{deWit:1975veh}
\begin{equation}
S_{\mu}(x) = 
-\frac{1}{2} F_{\rho\nu}^{a}(x) \sigma_{\rho\nu} \gamma_{\mu}
\lambda^{a}(x),
\end{equation}
whose divergence is proportional to the gluino mass,
\begin{equation}
\partial_{\mu} S_{\mu}(x) = m_{0} \chi(x),
\end{equation}
where
\begin{equation}
\chi(x) = \frac{1}{2} F_{\rho\nu}^{a}(x) \sigma_{\rho\nu} \lambda^{a}(x).
\end{equation}
Both $S_{\mu}(x)$ and $\chi(x)$ are spinorial quantities.

The corresponding formal SUSY Ward identities in the quantised theory
with a mass term are
\begin{equation}
\big\langle \partial_{\mu} S_{\mu}(x) Q(y) \big\rangle 
= m_0 \big\langle \chi(x) Q(y) \big\rangle 
- \bigg\langle \frac{\delta Q(y)}{\delta \bar{\epsilon}(x)} \bigg\rangle.
\end{equation}
Here $Q(y)$ is any suitable insertion operator, and the last term represents
a contact term given by the SUSY variation of $Q(y)$, which vanishes if
$Q(y)$ is localised at space-time points different from $x$.

A quantised theory is, however, only properly defined once it is regularised.
Regularisation on a lattice and renormalisation leads to significant
modifications of the Ward identities
\cite{Curci:1986sm,Farchioni:2001wx}. For details we refer to the cited
articles, and just report the main results.
In addition to the soft breaking by the gluino mass term, supersymmetry
is broken by the lattice regularisation. Analysis of the relevant
operators indicates that a continuum limit should exist
with the following characteristics. First, the gluino mass receives an
additive renormalisation, leading to a subtracted gluino mass $m_S$. 
Second, and more important, the supercurrent mixes
with another dimension 7/2 current, namely
\begin{equation}
T_{\mu}(x) 
= F_{\mu\nu}^{a}(x) \gamma_{\nu} \lambda^{a}(x).
\end{equation}
Based on suitably defined SUSY transformations on the lattice
\cite{Curci:1986sm,Taniguchi:1999fc}, 
the resulting SUSY Ward identity, omitting contact terms, reads
\begin{equation}
\label{WI}
Z_S \big\langle \big( \nabla_{\mu} S_{\mu}(x) \big) Q(y)\big\rangle 
+ Z_T \big\langle \big( \nabla_{\mu} T_{\mu}(x) \big) Q(y) \big\rangle
= m_S \big\langle \chi(x) Q(y) \big\rangle + O(a),
\end{equation}
where $Z_S$ and $Z_T$ are
renormalisation coefficients. A renormalised supercurrent can then be
defined through $S_{\mu}^{R} = Z_S S_{\mu} + Z_T T_{\mu}$.

In our numerical simulations we use a lattice action proposed by Curci and
Veneziano \cite{Curci:1986sm}, which is built in analogy to the Wilson
action of QCD for the gauge field and Wilson fermion action for the gluino.
Both supersymmetry and chiral symmetry are broken on the lattice, but they
are expected to be restored in the continuum limit if the gluino mass $m_S$
is tuned to zero. The Curci-Veneziano action for SYM theory on the lattice
is given by $S = S_g + S_f$, where
\begin{equation}
S_g = -\frac{\beta}{N_{c}} \sum_{p} \mathrm{Re\,Tr}\ U_{p}
\end{equation}
is the gauge field action with inverse gauge coupling $\beta = 2N_c/g^2$,
summed over the plaquettes $p$, and
\begin{equation}
S_f =
\frac{1}{2} \sum_{x} \left\{ \bar{\lambda}^{a}_{x} \lambda_{x}^{a} 
- \kappa \sum_{\mu = 1}^{4} \left[ \bar{\lambda}^{a}_{x +
\hat{\mu}} V_{ab, x \mu} (1 + \gamma_{\mu}) \lambda^{b}_{x}
+ \bar{\lambda}^{a}_{x} V^{T}_{ab, x \mu} (1 - \gamma_{\mu})
\lambda^{b}_{x + \hat{\mu}} \right] \right\}
\end{equation}
is the fermion action, where $V_{ab, x \mu} = 2\,\Tr\,(U_{x\mu}^\dagger T_a
U_{x\mu} T_b)$ is the gauge field variable in the adjoint representation
($T^a$ are the generators of SU($N_c$)), and the hopping parameter $\kappa$
is related to the bare gluino mass via $\kappa = 1/(2 m_{0} + 8)$.
In our numerical simulations the fermion action is $O(a)$ improved by
addition of the clover term with the one-loop coefficient specific for this
model \cite{Musberg:2013foa}.

The supercurrent $S_{\mu}(x)$ and the density $\chi(x)$ can be
defined on the lattice in various ways, differing by $O(a)$ terms. We choose
the local transcriptions of the continuum forms,
\begin{align}
S_{\mu}(x) &= - \frac{1}{2} P^{(cl)a}_{\rho\nu}(x)\,
\sigma_{\rho\nu} \gamma_{\mu} \lambda^{a}(x),\\
\chi(x) &= \frac{1}{2} P^{(cl)a}_{\rho\nu}(x)\, \sigma_{\rho\nu} \lambda^{a}(x),
\end{align}
which have led to the best signals in previous numerical studies. For this choice,
$\nabla_{\mu}$ indicates the symmetric lattice derivative, and
$P^{(cl)}_{\rho\nu}(x)$ is the clover plaquette.

The supersymmetric continuum limit is obtained at vanishing
gluino mass $m_S$. The value of the critical hopping parameter $\kappa_c$, where
$m_S$ is zero, has to be determined numerically. With suitable choices of
$Q(y)$, this can be achieved with the lattice SUSY Ward identity. The
expectation values appearing in Eq.~\eqref{WI} can be evaluated in the Monte
Carlo calculations. This allows to obtain the coefficient $m_S / Z_S$, which
in turn enables us to locate the point $m_S = 0$. An alternative tuning is
obtained from the signals of a restored chiral symmetry, see below. It is
expected that both are consistent up to lattice artefacts. The investigation
of the SUSY Ward identities allows to confirm this scenario and to estimate
the relevant lattice artefacts.

%%%%%%%%%%%%%%%%%%%%%%%%%%%%%%%%%%%%%%%%%%%%%%%%%%%%%%%%%%%%%%%%%%%%%%%%%%%%%%%%
\section{Numerical analysis of SUSY Ward identities}

In the numerical analysis it is convenient to project to zero momentum by
summing the operators over the three spatial coordinates. As a result one
obtains a Ward identity for each time slice separation $t = x_4 - y_4$. Each
term in Eq.~\eqref{WI} is a $4 \times 4$ matrix in Dirac space and can be
expanded in the basis of $16$ Dirac matrices. Using discrete symmetries one
can show that only two non-trivial independent equations
survive~\cite{Farchioni:2001wx}:
\begin{equation}
\label{WI2}
\begin{split}
\hat{x}_{1,t,1} + (Z_T Z^{-1}_S) \hat{x}_{1,t,2} 
&= (a m_S Z^{-1}_S) \hat{x}_{1,t,3}, \\
\hat{x}_{2,t,1} + (Z_T Z^{-1}_S) \hat{x}_{2,t,2} 
&= (a m_S Z^{-1}_S) \hat{x}_{2,t,3},
\end{split}
\end{equation}
where $O(a)$ terms are omitted, and
\begin{align}
\hat{x}_{1,t,1} &\equiv \sum_{\vec{x}} \big\langle \nabla_4 S_4(x) Q(0) \big\rangle,
&\hat{x}_{2,t,1} &\equiv \sum_{\vec{x}}\big\langle \nabla_4 S_4(x) \gamma_4 Q(0) 
\big\rangle,\nonumber\\
\hat{x}_{1,t,2} &\equiv \sum_{\vec{x}} \big\langle \nabla_4 T_4(x) Q(0) \big\rangle, 
&\hat{x}_{2,t,2} &\equiv \sum_{\vec{x}} \big\langle \nabla_4 T_4(x)\gamma_4 Q(0) 
\big\rangle,\\
\hat{x}_{1,t,3} &\equiv \sum_{\vec{x}} \big\langle \chi(x) Q(0) \big\rangle,
&\hat{x}_{2,t,3} &\equiv \sum_{\vec{x}} \big\langle \chi(x)\gamma_4 Q(0) 
\big\rangle \nonumber.
\end{align}
Here, traces over spinorial indices are implied.
Concerning the insertion operator, it turned out that
\begin{equation}
Q(y) = \chi^{(\mathrm{sp})}(y) = \sum_{i<j} \sigma_{ij} 
P^{(cl)a}_{ij}(y) \lambda^{a}(y), \qquad i,j \in \{1,2,3\}
\end{equation}
gives the best signal. The signal-to-noise ratio is improved further by applying
APE and Jacobi smearing to this operator.

The six different correlators $\hat{x}_{b,t,\alpha}$ are estimated numerically in
our Monte Carlo simulations for gauge group SU(3).
The usual estimators for these expectation values are the
numerical averages of the corresponding observables over the Monte Carlo run.
Let us call these averages $x_{b,t,\alpha}$.
They are random variables with expectation
values $\hat{x}_{b,t,\alpha} \equiv \langle x_{b,t,\alpha} \rangle$.
It should be noted that
only data at $t \geq 3$ are being considered in order to avoid contamination
by contact terms.

For each $t$ the two equations \eqref{WI2} could be solved for 
\begin{equation}
A = Z_T Z^{-1}_S \quad \text{and} \quad B = a m_S Z^{-1}_S.
\end{equation}
Taking all $t$ together, however, we have an overdetermined set of equations
for these two coefficients. The aim is to find solutions for $A$ and $B$
numerically such that with the measured values $x_{b,t,\alpha}$ the
equations are satisfied approximately in an optimal way. In previous studies
for gauge group SU(2) the coefficients $A$ and $B$ have been calculated by
means of a minimal chi-squared method, as proposed in
\cite{Farchioni:2001wx}. The correlators $x_{b,t,\alpha}$ are, however,
statistically correlated amongst each other, in particular for nearby values
of $t$, and these correlations have not been taken into account.

In order to improve on this point, we have developed a method, which takes all
correlations fully into account, so that more reliable results and error
estimates can be obtained. The approach is based on the method of
generalised least squares \cite{Marshall:2005}.

The equations \eqref{WI2} hold for the expectation values. With the notation
\begin{equation}
A_1 = 1,\quad A_2 = A,\quad A_3 = -B,
\end{equation}
and the double index $i=(b,t)$, they can be written
\begin{equation}
\sum_{\alpha} A_{\alpha} \, \hat{x}_{i\alpha} = 0.
\end{equation}

Let $C_{i\alpha,j\beta} = \langle x_{i\alpha} x_{j\beta} \rangle
- \langle x_{i\alpha} \rangle \langle x_{j\beta} \rangle$
be the covariance matrix of $x_{i\alpha}$.
The probability distribution of the $x_{i\alpha}$ is given by
$P \sim \exp (-L)$ with
\begin{equation}
L = \frac{1}{2} \sum_{i,\alpha,j,\beta} (x_{i\alpha} - \hat{x}_{i\alpha}) 
M_{i\alpha,j\beta} (x_{j\beta} - \hat{x}_{j\beta}),
\qquad M = C^{-1}.
\end{equation}
For estimating $A_\alpha$ we employ the method of maximum likelihood in the
following way.
\begin{enumerate}
\item
For given $x_{i\alpha}$, consider $A_{\alpha}$ to be fixed and determine
$\hat{x}_{i\alpha}$ such that $P$ is maximal under the constraint
$\sum_{\alpha} A_{\alpha} \, \hat{x}_{i\alpha} = 0$. The value
$P_{\text{max}}(A_\alpha)$ at maximum depends on $A_\alpha$.
\item
Find $A_\alpha$ such that $P_{\text{max}}(A_\alpha)$ is maximal.
\end{enumerate}
Minimising $L$ with the help of Lagrange multipliers gives
\begin{equation}
x_{i\alpha} - \hat{x}_{i\alpha} 
= \sum_{j,\beta} C_{i\alpha,j\beta} A_{\beta} \sum_{k\gamma} 
(D^{-1})_{jk} x_{k\gamma} A_{\gamma}
\end{equation}
and
\begin{equation}
L_{\text{min}} = \frac{1}{2} \sum_{i,\alpha,j,\beta} (A_{\alpha} x_{i\alpha})
(D^{-1})_{ij} (A_{\beta} x_{j\beta}),
\end{equation}
where
\begin{equation}
D_{ij} \doteq \sum_{\alpha,\beta} A_\alpha C_{i\alpha,j\beta} A_\beta .
\end{equation}
For given $A_{\alpha}$ the matrix $D_{ij}$ is estimated, up to an irrelevant
constant factor, from the measured values by
\begin{equation}
D_{ij} = \sum_{\alpha, \beta} A_{\alpha} A_{\beta}
\tilde{C}_{i\alpha,j\beta},
\end{equation}
where $\tilde{C}_{i\alpha,j\beta}$ is the covariance matrix of the
primary observables.

Now the minimum of $L_{\text{min}}(A_{\alpha})$ as a function of the
parameters $A_2$ and $A_3$ ($A_1=1$) has to be found. Because $D_{ij}$
depends on the $A_{\alpha}$, it is not possible to do this analytically, and
we determine the global minimum numerically, thus obtaining $A_2$ and $A_3$.
To get the statistical errors we re-sample the data and apply the jackknife
method, repeating the whole procedure for each jackknife sample.
In this way we arrive at our final result for $B = a m_S Z_S^{-1}$.

%%%%%%%%%%%%%%%%%%%%%%%%%%%%%%%%%%%%%%%%%%%%%%%%%%%%%%%%%%%%%%%%%%%%%%%%%%%%%%%%
\section{Results for SU(3) SYM}

For SYM theory with gauge group SU(3) we have applied the method to our
current simulation ensembles obtained with $O(a)$ improved clover fermion
action \cite{Ali:2018dnd} at different inverse gauge couplings $\beta$ and
hopping parameters $\kappa$. At two lattice spacings, corresponding to
$\beta = 5.4$ and $5.5$, the available statistics has allowed to obtain
reliable results for the Ward identities. From the results for the gluino
mass parameter $a m_S Z_S^{-1}$ the value of $\kappa_c$, where $m_S$
vanishes, can be estimated.

Comparing the results for $a m_S Z_S^{-1}$ with those from the earlier
method, which does not properly take the correlations into account, we
find that the values are compatible within errors, but this time
we have a precise and reliable estimate of the errors. As examples,
the results of both methods for $\beta = 5.5$ are shown in 
Tab.~\ref{tab:comparison}
\begin{table}[ht!]
\begin{center}
\begin{tabular}{|l|l|l|l|l|l|l|l|}
\hline
\hspace{15pt}$\kappa$ & 0.1637 & 0.1649 & 0.1667 & 0.1673 & 0.1678 & 0.1680 
& 0.1683 \\
\hline
previous & 0.489(26) & 0.343(7) & 0.176(4) & 0.123(3) & 0.081(3) & 0.057(4) 
& 0.025(4) \\
GLS & 0.494(42) & 0.348(8) & 0.178(4) & 0.123(3) & 0.081(2) & 0.056(5) 
& 0.024(6) \\
\hline
\end{tabular}
\caption{Results for $a m_S Z_S^{-1}$ from the previous method and from the
generalised least squares (GLS) method for our ensembles at $\beta = 5.5$.}
\label{tab:comparison}
\end{center}
\end{table}

An alternative way to estimate $\kappa_c$ in the Monte Carlo calculations
employs the mass of the adjoint pion $\api$, see e.\,g.\
\cite{Demmouche:2010sf}. The $\api$ is an unphysical particle in SYM theory.
However, by arguments based on the OZI-approximation
\cite{Veneziano:1982ah}, and in the framework of partially quenched chiral
perturbation theory~\cite{Munster:2014cja}, the squared mass $m_{\api}^2$ is
expected to vanish linearly with the gluino mass close to the chiral limit.

\begin{figure}[t]
\centering
\includegraphics[width=0.49\textwidth]{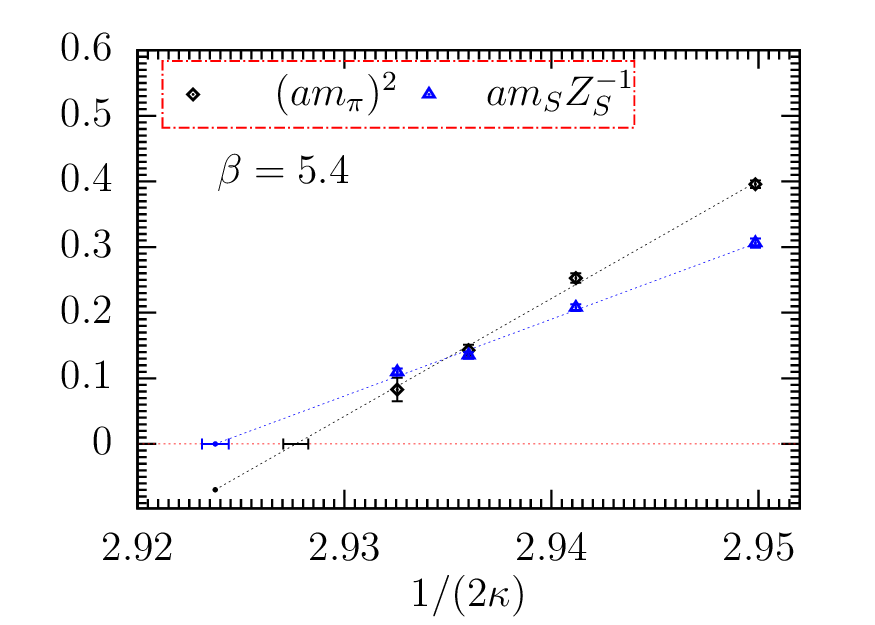}
\includegraphics[width=0.49\textwidth]{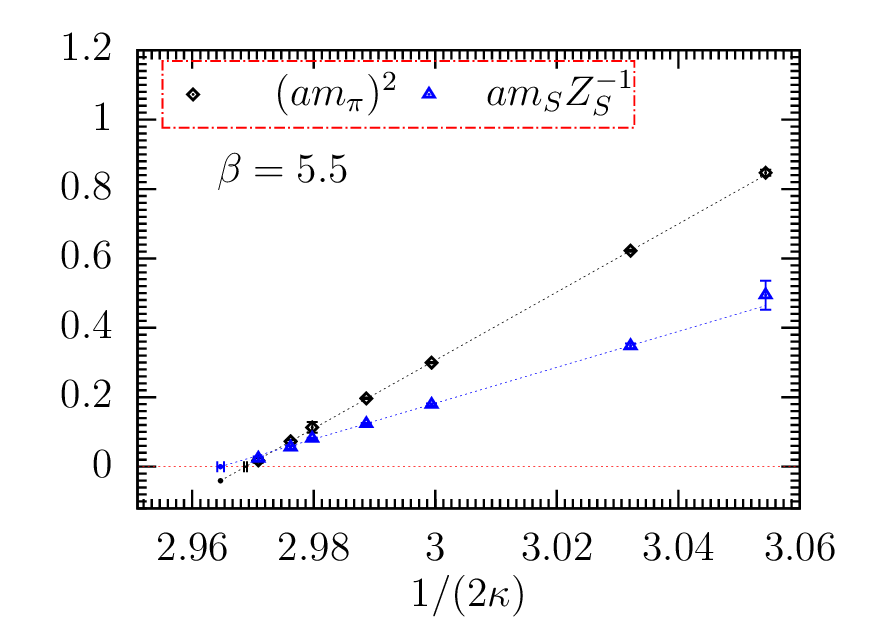}
\caption{The subtracted gluino mass $a m_S Z_S^{-1}$ and the squared
adjoint pion mass $(a m_{\api})^2$ as
a function of $1/(2\kappa)$, and the corresponding
extrapolations towards the chiral point ($\kappa_c$) for two values
of $\beta$.}
\label{Kc}
\end{figure}

In Fig.~\ref{Kc} we show $a m_S Z_S^{-1}$ and $(a m_{\api})^2$ as a function of
$1/(2\kappa)$ for our two values of $\beta$. Both quantities depend linearly
on $\kappa^{-1}$ within errors, as expected, and yield independent estimates
of the value of $\kappa_c$.

The values of $\kappa_c$ obtained from the Ward identities and from
$m_{\api}^2$ are very close to each other, but there is a small difference.
This discrepancy should be due to lattice artefacts, and we expect it to
disappear in the continuum limit.

In the case of lattice QCD, Wilson chiral perturbation theory to leading
order shows a shift linear in $a$ in the dependence of the squared pion mass
on the quark mass:
\begin{equation}
m_{\pi,\text{LO}}^2 = 2 B_0 m_{q} + 2 W_0 a\,,
\end{equation}
with certain low-energy constants $B_0$ and
$W_0$~\cite{Sharpe:1998xm,Rupak:2002sm}. On the other hand, for the PCAC
quark mass, defined by means of the chiral Ward identity, exactly the same
shift is present in leading order~\cite{Sharpe:2004ny},
\begin{equation}
2 B_0 m_{\text{PCAC,LO}} = 2 B_0 m_{q} + 2 W_0 a\,.
\end{equation}
Consequently, at vanishing pion mass, the remnant $m_{\text{PCAC}}$ is of
order $a^2$, and this result is not changed in higher orders of chiral
perturbation theory,
\begin{equation}
m_{\text{PCAC}} = O(a^2) \quad \text{at} \quad m_{\pi}^2 = 0.
\end{equation}

\begin{figure}[t]
\centering
\includegraphics[width=0.49\textwidth]{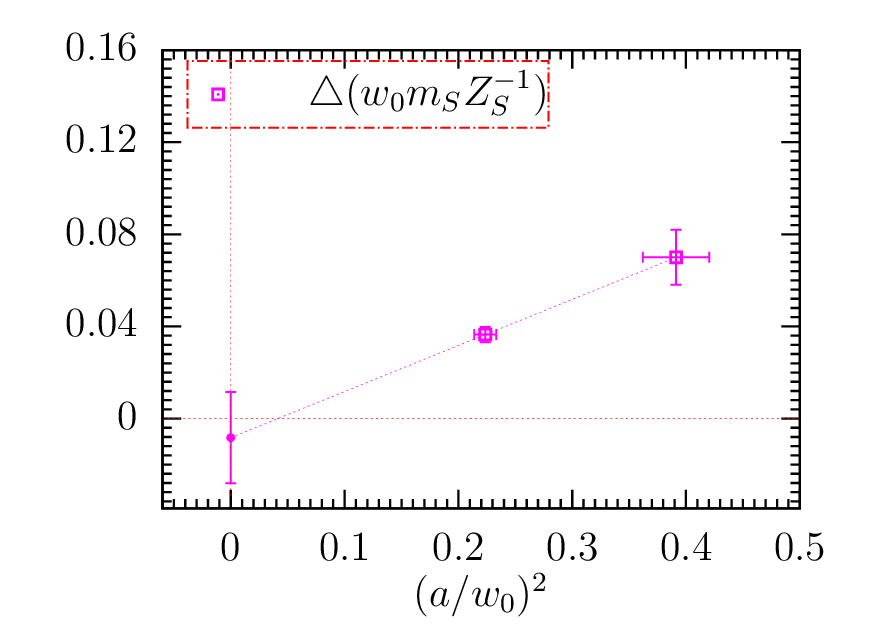}
\caption{The remnant gluino mass $\Delta (w_0 m_S Z_S^{-1})$ 
at vanishing $m_{\api}^2$ as a function
of the lattice spacing squared. The common scale is set through the gradient
flow parameter $w_0$.}
\label{Deltam}
\end{figure}

In SYM the adjoint pion mass can be calculated in partially quenched chiral
perturbation theory~\cite{Munster:2014cja}. We haven't evaluated the
contributions from the lattice terms explicitly, but the structure of terms
is similar to those for
QCD, and therefore we expect that in SYM the remnant gluino mass $\Delta
m_S$ at vanishing adjoint pion mass is of order $a^2$, too. In order to
check this numerically, the masses have to be expressed in a physical scale.
We use the scale $w_0$, defined through the gradient flow; for details see
\cite{Ali:2018dnd}. In Fig.~\ref{Deltam} we show the remnant gluino mass as
a function of the squared lattice spacing $a^2$. The line through the points
extrapolates to zero within errors. For an analogous plot linear in
$a$ this is by far not the case. Having only two points available, one has
to be cautious drawing conclusions, but the result clearly indicates that the
remnant gluino mass
$\Delta m_S$ vanishes proportional to $a^2$ in the continuum limit.

%

%%%%%%%%%%%%%%%%%%%%%%%%%%%%%%%%%%%%%%%%%%%%%%%%%%%%%%%%%%%%%%%%%%%%%%%%%%%%%%%
\section{Conclusions}

We have presented a method for the numerical analysis of SUSY Ward
identities in supersymmetric Yang-Mills theory on a lattice, which employs
the expectation values of the relevant operators on a range of time slices.
The statistical correlations between all observables are taken into account
by means of a generalised least squares procedure. Applied to SUSY
Yang-Mills theory with gauge group SU(3), the value of the hopping
parameter, where the renormalised gluino mass vanishes, can be estimated,
and is in rough agreement with the estimation using the adjoint pion mass.
The difference between the estimates appears to vanish in the continuum
limit. Our results represent the first continuum extrapolation of SUSY Ward
identities. The scaling of lattice artefacts as of $O(a^2)$ is in agreement
with theoretical expectations.

%%%%%%%%%%%%%%%%%%%%%%%%%%%%%%%%%%%%%%%%%%%%%%%%%%%%%%%%%%%%%%%%%%%%%%%%%%%%%%%%
\section*{Acknowledgments}

The authors gratefully acknowledge the Gauss Centre for Supercomputing
e.~V.\,\linebreak(www.gauss-centre.eu) for funding this project by providing
computing time on the GCS Supercomputer JUQUEEN and JURECA at J\"ulich
Supercomputing Centre (JSC) and SuperMUC at Leibniz Supercomputing Centre
(LRZ). Further computing time has been provided on the compute cluster PALMA
of the University of M\"unster. This work is supported by the Deutsche
Forschungsgemeinschaft (DFG) through the Research Training Group ``GRK 2149:
Strong and Weak Interactions - from Hadrons to Dark Matter''. G.B.
acknowledges support from the Deutsche Forschungsgemeinschaft (DFG) Grant
No.~BE 5942/2-1.

%%%%%%%%%%%%%%%%%%%%%%%%%%%%%%%%%%%%%%%%%%%%%%%%%%%%%%%%%%%%%%%%%%%%%%%%%%%%%%%%

%%%%%%%%%%%%%%%%%%%%%%%%%%%%%%%%%%%%%%%%%%%%%%%%%%%%%%%%%%%%%%%%%%%%%%%%%%%%%%%%
\end{document}